\documentclass[aps,twocolumn,groupeaddress,prb]{revtex4}

\usepackage{amsfonts, amsmath, amssymb,latexsym}
\usepackage{subfigure}
\usepackage{graphicx}

\newcommand{\lab}{\left<}
\newcommand{\rab}{\right>}

\newcommand{\exmu}{\mu^{\mathrm{ex}}}  
\newcommand{\ep}{\varepsilon}        

\newcommand{\me}{\mathrm{e}}

\newcommand{\dif}{\mathrm{d}}

\begin{document}

\title{Quasi-chemical theory with a soft cutoff}

\author{Shaji Chempath}
\affiliation{Theoretical Division, Los Alamos National Laboratory, Los
Alamos, NM 87545 USA}
\author{Lawrence R. Pratt}
\affiliation{Department of Chemical and Biomolecular Engineering, Tulane
University, New Orleans, LA 70118}\email{lpratt@tulane.edu} 
\author{M. E. Paulaitis} 
\affiliation{Department of Chemical and Biomolecular Engineering, Ohio
State University, Columbus OH 43210 USA}

\date{\today}
\begin{abstract} In view of the wide success of molecular quasi-chemical
theory of liquids, this paper develops the  soft-cutoff version of that
theory. This development has important practical consequences in the
common cases that the packing contribution dominates the solvation free
energy of realistically-modeled molecules because treatment of hard-core
interactions usually requires special purpose simulation methods.  In
contrast, treatment of smooth repulsive interactions is typically
straightforward on the basis of widely available software.  This
development also shows how fluids composed of molecules with smooth
repulsive interactions can be treated analogously to the molecular-field
theory of the hard-sphere fluid. In the treatment of liquid water,
quasi-chemical theory with soft-cutoff conditioning doesn't  change the
fundamental convergence characteristics of the  theory using hard-cutoff
conditioning. In fact, hard cutoffs are found here to work better than
softer ones.
\end{abstract}


\maketitle

\section{Introduction}

Recent developments of quasi-chemical theory constitute a fresh attack
on the theory of solutions from a molecular scale, a point that has been
noted before.\cite{PrattLR:Quatst,Paulaitis:02,beck2006,CPMS}
Quasi-chemical theory provides a natural organization for calculations
of ion-water chemical interactions in the treatment of ion hydration
free energies.\cite{Rempe:JACS:2000,AsthagiriD:Hydsaf,CPMS} A remarkably
natural quasi-chemical theory for the statistical thermodynamics of
liquid water\cite{shah:144508} can be implemented on the basis of the
massive data sets of molecular simulations and provides a realistic
direct evaluation of the entropy of that liquid.  For non-polar solutes
dissolved in liquid water, quasi-chemical theory avoids physical
assumptions of the van der Waals (perturbative) type that are otherwise
common, and thus resolves questions centering on reference
system assumptions.\cite{AsthagiriD.:NonWth,Asthagiri:2008p1418} As originally
developed with a sharp cutoff to spatially define the inner shell,
quasi-chemical theory implemented as a molecule-field theory for the
hard-sphere fluid provides an equation of state as accurate as the most
accurate previous theory.\cite{PrattLawrenceR.:Selmft}

This quasi-chemical theory can be expressed entirely as statistical
modeling of the distribution $P(\ep) $ of energies $\ep$ binding a
distinguished molecule to the medium.  A goal is to discriminate effects of
close, or \emph{inner-shell}, neighbors of that molecule from the
effects of the more distant material.   This is done by formulating a
statistical condition to identify inner-shell molecules.   We will
express our condition   with $\chi$, a function
indicating that the inner shell is \emph{empty} when $\chi = 1$. $\chi$ has been
assumed to be either 0 or 1, sharply changing between those
possibilities.  We then direct attention to the conditional probability
$P(\ep \vert \chi = 1) $ in order to define an outer-shell contribution,
and then further use chemical concepts to evaluate the remainder
associated with occupancy of the inner-shell.

This quasi-chemical theory with sharp- or hard-cutoff conditioning works
satisfactorily both for the free energy of liquid
water\cite{shah:144508} and for the hydration free energy of
CF$_4$(aq).\cite{AsthagiriD.:NonWth} In the latter case, the theory
generates a result similar to a van der Waals theory without requiring
those assumptions in advance.   A prominent contribution to that result
is a hard-core or \emph{packing} contribution associated with the sharp
$\chi = 1$ condition.   On an intuitive basis, it is widely asked
whether a more natural packing contribution might be based upon smooth
repulsive interactions.  In the contrasting case of liquid water, more
aggressive conditioning is required, and the packing contribution
clearly does not have the logical status of a contribution of
repulsive-force reference interaction considered to be similar to actual
intermolecular interactions.   The present work is designed to clarify
these contrasting points.

We anticipate the results below by noting here that the theory for the
soft-cutoff conditioning can be developed in straightforward analogy
with that for hard-cutoff conditioning. This development has important
practical consequences because a packing contribution frequently
dominates the solvation free energy of realistically-modeled molecules but
treatment of hard-core interactions associated with a hard cutoff
usually requires special purpose calculations.  Examples of the special
considerations that apply to simulation of liquids of hard-core-model
polyatomic molecules can be found in the
references.\cite{Stratt:1981p3179,Smith:1997p3228}  In contrast,
treatments of smooth repulsive interactions are typically
straightforward on the basis of widely available software.  The present
development implies also that fluids composed of molecules with smooth
repulsive interactions can be treated by analogy with the
molecular-field theory of the hard-sphere
fluid.\cite{PrattLawrenceR.:Selmft}

In a case such as CF$_4$(aq) where only minimal conditioning is
required,\cite{AsthagiriD.:NonWth} we propose below a soft cutoff that
should reduce the variance of estimates of the required contributions.
In the case of liquid water where aggressive conditioning is
required,\cite{shah:144508} we can systematically vary the softness of
the cutoff defining the inner shell, and attempt to learn what works
best.

\section{Theory}
Quasi-chemical theory with sharp- or hard-cutoff conditioning,
which we aim to generalize, can be expressed as
\begin{subequations}
\begin{equation}
\beta\exmu  =  -  \ln  \lab\lab \chi \rab\rab_{0}   + \ln  \lab \chi \rab   
+ \ln\int \me^{ \beta \ep }P(\ep | \chi =1) \dif\ep ~, \label{eqn:fnrg0}
\end{equation}
\begin{multline}
\beta\exmu  \approx   -  \ln \lab\lab \chi \rab\rab_{0} + \ln \lab \chi \rab   
      \\ + \beta\lab \varepsilon | \chi=1 \rab 
      	+ \frac{\beta^2}{2}\lab \delta \varepsilon^2 | \chi=1 \rab ~.  
\label{eqn:fnrg1}
\end{multline}
\end{subequations}
The notation $\lab\lab \ldots \rab\rab_{0}$ indicates the  average over
the thermal motion of the system \emph{and} the solute with no
interaction between them, thus the subscript 0.  $\chi$ is an indicator
function and  $\chi=1$ defines our condition.  As an example, for the
application to liquid water  $\chi=1$ when there are no O-atoms of bath
molecules within a radius $\lambda$ from the O-atom of a distinguished
water molecule. Then $\chi=0$ if any solvent molecule \emph{is} present in that
inner-shell. The conditional averages indicate that statistics are
collected only when $\chi=1$. Significant conditioning produces slightly
sub-gaussian binding energy probability densities $P(\ep|\chi=1)$.
Slight sub-gaussian behavior leads  to errors of
1 kcal/mol in the  excess chemical
potentials calculated with Eq.~\eqref{eqn:fnrg1}.\cite{shah:144508,Crete2008} Nevertheless, Eq.~\eqref{eqn:fnrg1}
adopts the gaussian approximation. 

The generalization that we seek can be economically identified on the
basis of a few primitive relations.  The first is the \emph{rule of
averages}
\cite{CPMS}
\begin{equation}
\label{eqn:ruleofavgs}
\lab F \rab =  
\frac{
\left\langle\left\langle \me^{-\beta \Delta U } F \right\rangle\right\rangle_0
}{
\left\langle\left\langle \me^{-\beta\Delta U}\right\rangle\right\rangle_0
}
 = \me^{\beta\mu^{\mathrm{ex}}}\left\langle\left\langle \me^{-\beta \Delta U } F \right\rangle\right\rangle_0~.
\end{equation}
Then, 
\begin{equation}
\label{eqn:pdt0}
\lab \me^{\beta\Delta U} \chi \rab
=\me^{\beta\mu^{\mathrm{ex}}}
\left\langle\left\langle\chi\right\rangle\right\rangle_0
\end{equation}
is an example of specific utility for our argument.
For an indicator function $\chi$ we have in addition
\begin{equation}
\label{eqn:chimult}
\lab G  \chi \rab = \lab G  | \chi=1 \rab  \lab \chi \rab  ~,
\end{equation}
and $\lab \chi \rab$ is a probability.
Collecting these relations yields
\begin{equation}
\label{eqn:pdt}
\me^{\beta\mu^{\mathrm{ex}}} = \lab \me^{\beta\Delta U} | \chi=1 \rab \times 
\left( \frac{\lab\chi\rab}{\lab\left\langle\chi\right\rangle\rab_0} \right)~, 
\end{equation}
or  Eq.~\eqref{eqn:fnrg0} after evaluating the logarithm. For a sharp
cutoff case, this is obtained directly from the standard thermodynamic
difference formula.\cite{shah:144508,Crete2008} Beyond traditional statistical
thermodynamics, however, this formula only depends on the fact that
$\chi$ is an indicator function.

We obtain soft-cutoff conditioning by designing an indicator function  which
categorizes a solvent molecule at a particular configuration to be either
{\bf{in}} or {\bf{out}} of the inner shell according to a statistical rule. For example,
let the designated or  solute molecule be at a fixed position, and
consider the  $i^{th}$ solvent molecule again at a definite
configuration denoted by $\mathcal{R}_i$.  We take $0\le
b(\mathcal{R}_i)\le 1$ to be the probability that the $i^{th}$ solvent
molecule is {\bf{in}}, independent of any other characteristic.  [For
the cases considered above $b(\mathcal{R}_i)$ only depends on the
three-dimensional position of the O (oxygen) atom of the $i^{th}$ water molecule.]
For each configuration, we could then sample $b(\mathcal{R}_i)$ for each
solvent molecule, and thereby identify each solvent molecule as either
{\bf{in}} or {\bf{out}}.  Exhaustive \textbf{in}-\textbf{out} sampling
at a fixed configuration produces
\begin{equation}
\label{eqn:softchi}
\bar{\chi} = \prod_{j}(1-b(\mathcal{R}_j))
\end{equation} 
as a sampling-averaged indicator function. The formula specifically
indicating a  soft-cutoff is then
\begin{equation}
\label{eqn:softcut2}
\me^{\beta\mu^{\mathrm{ex}}} = \lab \me^{\beta\Delta U} | \chi=1 \rab \times 
\left( \frac{\lab\bar\chi\rab}{
\lab\left\langle\bar\chi\right\rangle\rab_0
} \right)~. 
\end{equation}
The hard-cutoff theory is recovered if step functions are chosen for the
$b(\mathcal{R}_i)$.

It is interesting to note that Eq.~\eqref{eqn:pdt0} is specialization of 
\begin{equation}
\label{eqn:softcut1}
\me^{\beta\mu^{\mathrm{ex}}} = 
\frac {\lab \me^{\beta\Delta U} W \rab  }{\lab\left\langle W\right\rangle\rab_0} ~,
\end{equation} 
to the case that $W = \bar\chi $ is an indicator function.
Bennett\cite{BENNETTCH:Effefe} has studied such forms without the
restriction that $W$ be an indicator function, and found  $W \propto (1+
z \me^{\beta\Delta U})^{-1}$ to be  statistically optimal with a
particular value of $z$. In contrast, our goal is the
dis-entanglement of the  effects of longer-ranged intermolecular
interactions from   packing and chemical contributions, so we do not
follow Bennett's results here. A potential advantage of
Eq.~\eqref{eqn:softcut2} is that $\lab\bar\chi\rab$ is an occupancy probability and
therefore can be modeled on a pattern of chemical
equilibria, as we elaborate below.  

Nevertheless, if minimal conditioning is all that is required, then
a straightforward suggestion for choice of the 
indicator function is 
$\bar\chi = \exp\left(-\beta u_0\right)$
for a smooth repulsive molecular-pair interaction, $u_0$,
that matches $\Delta U$ at short range. In view of Eq.~\eqref{eqn:softcut1}, this choice would
reduce the variance of the numerator of Eq.~\eqref{eqn:softcut2}.  Then
the denominator is 
$\lab\left\langle\bar\chi\right\rangle\rab_0 = 
\exp\left(-\beta \mu^{\mathrm{ex}}_0\right)$.

\subsection*{Chemical interpretation}
A general derivation of quasi-chemical theory \cite{Beck:2006} 
can be based upon an inclusion-exclusion development initiated
from the tautology
\begin{eqnarray}
1 = \prod_{j}\left\lbrack b(\mathcal{R}_j) + (1-b(\mathcal{R}_j))\right\rbrack~.
\label{eq:roo}
\end{eqnarray}
When included within the averaging brackets that 
yield the chemical potential, Eq.~\ref{eq:roo} leads to the general development \cite{Beck:2006} 
\begin{eqnarray}
\me^{-\beta\mu_\alpha^\mathrm{ex}}  =  \left( {1+\sum\limits_{n\ge 1} {K_n\rho^n}} \right)  
\times \left\langle {\left\langle
\me^{-\beta\Delta U_\alpha}\bar{\chi} \right\rangle }
\right\rangle _0~.
\label{qca2}
\end{eqnarray}
Since 
\begin{eqnarray}
\frac{\left\langle \left\langle\me^{-\beta\Delta U_\alpha}\bar{\chi} \right\rangle\right\rangle _0}
	{\left\langle \left\langle \me^{-\beta\Delta U_\alpha} \right\rangle\right\rangle _0} = \left\langle \bar{\chi}\right\rangle~,
\end{eqnarray}
Eq.~\eqref{qca2} implies
\begin{eqnarray}
\label{qca3}
\left\langle \bar{\chi}\right\rangle = 
\frac{1}{ {1+\sum\limits_{n\ge 1} {K_n\rho^n}}}~.
\end{eqnarray}
The $K_n$'s involve the cutoff function in  a natural
way.\cite{Beck:2006}   
Together with
\begin{eqnarray}
p( n) = \left\langle {N \choose n}
\left\lbrack\prod\limits_{j=1}^{j=n} b(\mathcal{R}_j)\right\rbrack
\left\lbrack\prod\limits_{k=n+1}^{k=N} \left(1-b(\mathcal{R}_k)\right)\right\rbrack
\right\rangle,
\label{eq:pndef}
\end{eqnarray}
 $ K_n
\rho^n$ is the ratio of the probability that the distinguished species has
$n$ ligands to the probability that it has zero (0) ligands
\begin{eqnarray}
\frac{p( n)}{p( 0) } 
= K_n\rho^n~.
\label{eq:defKa}
\end{eqnarray}
This is naturally interpreted on the basis  of the chemical equilibrium
\begin{eqnarray}
\mathrm{W}^* + n \mathrm{W} \leftrightharpoons  \mathrm{W^*W}_n 
\end{eqnarray}
where $\mathrm{W^*}$ denotes a water molecule with zero (0) ligands and
$\mathrm{W^*W}_n$ denotes that water molecule with $n$ ligands occupying
the inner shell. Given
Eq.~\eqref{qca3}, the probability that the distinguished molecule has
zero (0) ligands is $\left\langle \bar{\chi}\right\rangle$.  This establishes the thermodynamic significance of
that normalization factor. These interpretations are straightforward for
the sharp-cutoff case.  If the
probabilities of Eq.~\eqref{eq:defKa} are given by a Poisson distribution,
then $K_n = v^n/n!$ with $v$ the volume of the stencil.
For a specified function $b(\mathcal{R}_i)$, 
$p(n)$  and $K_n$ can be observed in conventional simulations where
the full physical solute-solvent interactions are expressed.

\begin{figure}[h]
\includegraphics[width=3.2in]{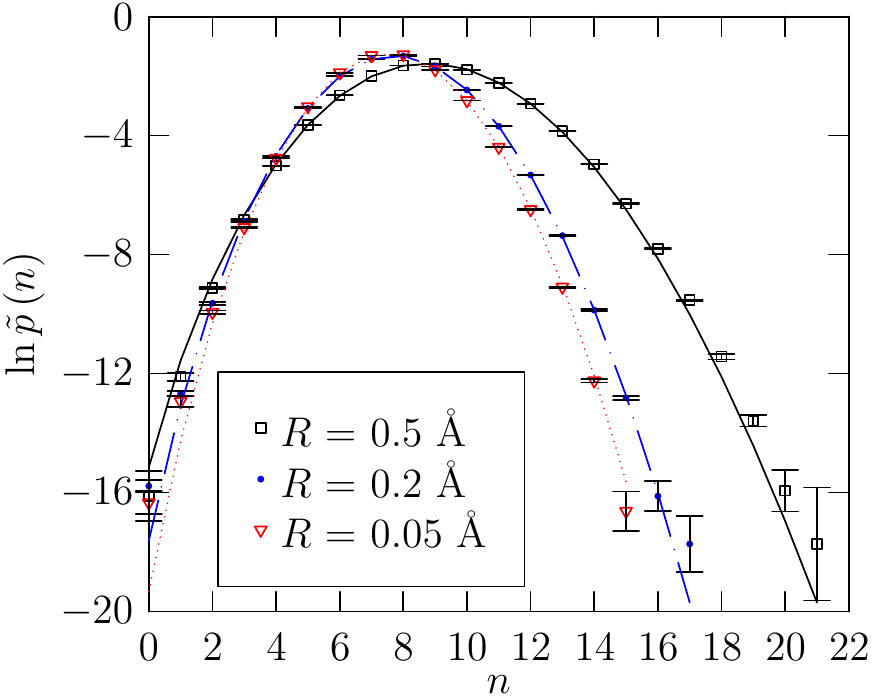}
\caption{For TIP3P water at $T$=300K and $p$=1atm: $\tilde
p\left(n\right)$ corresponding to the soft-cutoff function
of Eq.~\eqref{eqn:soft-cutoff} with $\lambda$ = 3.8\AA.  The smooth curves plotted are
two moment maximum entropy models obtained on the basis of the Poisson
default model.\cite{AsthagiriD.:NonWth}\label{fig:pn}}
\end{figure}

These analogies go further. For example, consider the packing
contribution $\left\langle {\left\langle\bar{\chi} \right\rangle }
\right\rangle _0$,\cite{Beck:2006}   and apply Eq.~\eqref{qca2} to the
case $\Delta U_\alpha$ = 0.  This yields
\begin{eqnarray}
1  =  \left( {1+\sum\limits_{n\ge 1} {\tilde{K}_n\rho^n}} \right)  
\times \left\langle {\left\langle\bar{\chi} \right\rangle }
\right\rangle _0~,
\label{eq:rawpacking00}
\end{eqnarray}
or
\begin{eqnarray}
\left\langle {\left\langle\bar{\chi} \right\rangle }
\right\rangle _0  =  \frac{1}{\left( {1+\sum\limits_{n\ge 1} {\tilde{K}_n\rho^n}} \right)},
\label{eq:rawpacking01}
\end{eqnarray}
with
\begin{eqnarray}
\frac{\tilde{p}(n)}{\tilde{p}(0)} =\tilde{K}_n \rho^n~.
\label{eq:defK}
\end{eqnarray}
Here the tildes indicate quantities observed \underline{\emph{without}}
the solute physically present, but with a soft stencil defined by the
indicator function $\bar\chi$.  Specifically
\begin{multline}
\tilde p( n) = \\
\left \langle \left\langle {N \choose n}
\left\lbrack\prod\limits_{j=1}^{j=n} b(\mathcal{R}_j)\right\rbrack
\left\lbrack\prod\limits_{k=n+1}^{k=N} \left(1-b(\mathcal{R}_k)\right)\right\rbrack
\right\rangle\right\rangle_0,
\end{multline}
 \emph{i.e.,} Eq.~\eqref{eq:pndef} for the present uncoupled case. 
 Fig.~\ref{fig:pn} gives example results of $\tilde p\left(n\right)$ for
 liquid water, and shows that soft-cutoff conditioning produces
 integer-referenced probabilities.

If we take $\bar\chi = \exp\left(-\beta u_0\right)$ as suggested above, 
Eq.~\eqref{eq:rawpacking01} gives the new and surprising result
\begin{eqnarray}
\mathrm{e}^{\beta \mu^{\mathrm{ex}}_0} = {1+\sum\limits_{n\ge 1} {\tilde{K}_n\rho^n}}
\label{eq:newANDsurprizing}\end{eqnarray}
for the free energies of model liquids composed of molecules with soft
repulsive interactions.   Eq.~\eqref{eq:newANDsurprizing} is, however,
directly analogous to the quasi-chemical theory for hard-core
fluids.\cite{PrattLR:Quatst,PrattLawrenceR.:Selmft} The utility of
Eq.~\eqref{eq:newANDsurprizing} is due, in part,  to the fact the
low-density limiting values $\lim_{\rho\rightarrow 0}\tilde{K}_n$ =
$\tilde{K}_n^{(0)}(T)$ can be evaluated once-and-for-all, and they
provide a remarkably natural initial model for that equation of
state.\cite{PrattLR:Quatst,PrattLawrenceR.:Selmft} Collecting preceding
results, Eq.~\eqref{eqn:softcut2} can be expressed in the appealing form
\begin{eqnarray}
\me^{\beta \mu^{\mathrm{ex}}} = \left\langle \me^{\beta \Delta U}\vert \chi=1\right\rangle
\times 
\left(
\frac{1+\sum\limits_{n\ge 1} \tilde{K}_n\rho^n}
{1+\sum\limits_{n\ge 1}K_n\rho^n}
\right)~.
\label{eq:soft-cutoff-ratio}
\end{eqnarray}

The developments of this subsection are expected to have the practical
consequence that the whole of theory of Ref.~8 
can be consistently implemented on the molecular dynamics simulations as
contrasted with Monte Carlo calculations.

Though results such as Eq.~\eqref{eq:newANDsurprizing} give a
compelling physical organization to the statistical problem of evaluating
$\beta \mu^{\mathrm{ex}}_0$, we emphasize that this approach is thoroughly tautological.
For example, the identification Eq.~\eqref{eq:defK} transforms 
Eq.~\eqref{eq:newANDsurprizing} to 
\begin{eqnarray}
\mathrm{e}^{\beta \mu^{\mathrm{ex}}_0} = \sum\limits_{n\ge 0} \frac{\tilde p( n)}{\tilde p(0)}
=\frac{1}{\tilde p(0)}
\end{eqnarray}
which is transparently correct.

\section{Computational Implementation}

All calculations were performed with the MUSIC
software \cite{url_music,gupta001} which is a set of FORTRAN 90 modules
for MC and MD simulations. A cubic simulation cell of edge-length 30
\AA\ was used in all NVT Monte Carlo simulations. Water was modeled with
the TIP3P forcefield.\cite{mahoney2000} Based on the density of TIP3P
water at 300K and 1 atm,\cite{Paschek:2004} 888 water molecules were
used in the simulations. All Lennard-Jones and electrostatic
interactions were cutoff at 9.0 \AA. The electrostatic interactions were
described by the shifted force method of Fennell and
Gezelter.\cite{fennell2006} 

\begin{figure}[h]
\includegraphics[width=3.2in]{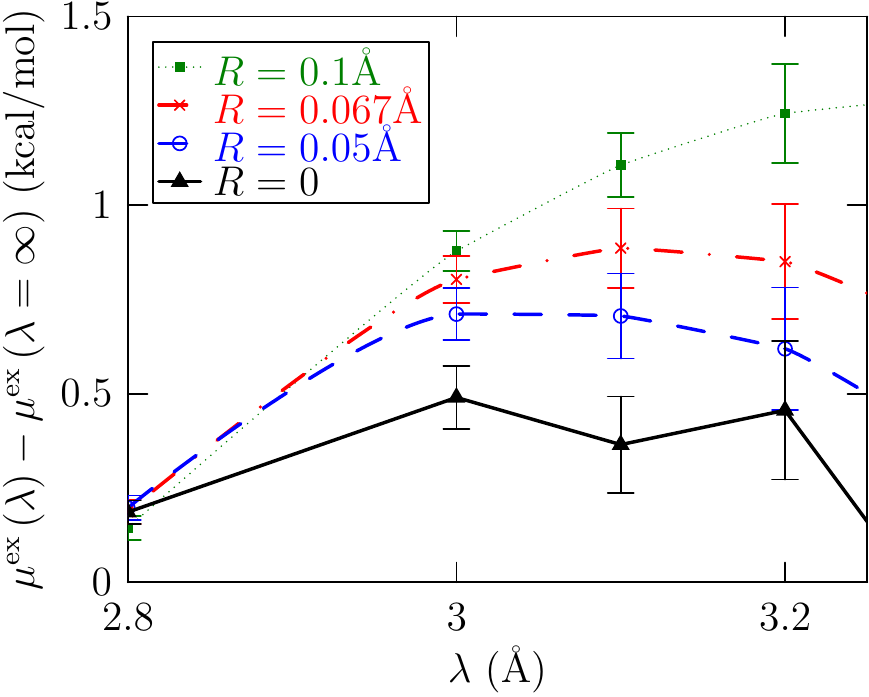}
\caption{The free energy discrepancy  on the basis of
Eq.~\eqref{eqn:fnrg1} for TIP3P water at 300K and normal density
utilizing different smooth cutoff functions of the form
Eq.~\eqref{eqn:soft-cutoff}.  \label{fig:softcut}}
\end{figure}

From a simulation record, the excess free energy can be directly evaluated by
applying Eq.~\eqref{eqn:softcut1}. Specifically for the soft
cutoff, averages such as $\left\langle \bar\chi\right\rangle$, 
$\left\langle \varepsilon \bar\chi \right\rangle$ = $\left\langle \varepsilon \left. \right|\chi =1 \right\rangle \left\langle \bar\chi\right\rangle$, and so on,
can be evaluated straightforwardly.   These were obtained
for the smooth cutoff function
\begin{equation}
1-b(r) = \frac{1}{1+\me^{-(r-\lambda)/R}}~,
\label{eqn:soft-cutoff}
\end{equation}
with $R$ a softness parameter.  The results of applying soft-cutoff
for water at 300K are shown in FIG.~\ref{fig:softcut}. 
The results there are all qualitatively similar in
overshooting the numerically exact value to which they must return with
$\lambda\rightarrow\infty$.  That overshoot is an indication that the
gaussian model places slightly too much weight in the high (unfavorable)
$\ep$ tail,\cite{shah:144508,Crete2008} \emph{i.e.,} the actual distribution
 is sub-gaussian.  
 
With the present generalized theory in hand, we investigated 
the following idea:   Consider a family of cutoff functions for
which the softness can be systematically varied.   An example is
\begin{eqnarray}
1 - b_m(r) = \me^{-\left(\lambda/r\right)^m}
\end{eqnarray}
with a parameter $m>3$.  With increasing $m$ this cutoff becomes
increasingly sharp.  A natural idea then is to choose $m$ to minimize
the variance of the conditional distribution $P(\ep | \chi =1)$
appearing in Eq.~\eqref{eqn:fnrg0}.   The goal is to narrow the
distribution so that deviations from the gaussian model for
$\mu^{\mathrm{ex}}$ (Eq.~\eqref{eqn:fnrg1}) are less important. We found
that the variance decreases with increasing $m$ (increasing sharpness)
towards an $m \sim \infty$ limiting value, the hard-cutoff case as a
practical matter. Hard-cutoff conditioning is always preferred 
for such cases.  The conclusion is that for systems and
calculations such as these, the results are always improved the more
that the conditioning eliminates statistical modeling of short-ranged
interactions, and hard cutoffs do that best.  Those interactions include
both associative and repulsive interactions that encompass both
chemical and packing contributions. This conclusion is clearly
consistent with the results shown in FIG.~\ref{fig:softcut}.

\section{Conclusions}

Quasi-chemical theory with a soft-cutoff can be developed in
straightforward analogy with the hard-cutoff theory. This development
has important practical consequences in the common cases that the
packing contribution dominates the solvation free energy of
realistically-modeled molecules because treatment of hard-core
interactions usually require special purpose simulation methods.  In contrast,
treatment of smooth repulsive interactions is typically straightforward
on the basis of widely available software.  This development also shows
how fluids composed of molecules with smooth repulsive interactions can
be treated analogously to the molecular-field theory of the hard-sphere
fluid.\cite{PrattLawrenceR.:Selmft}

In the treatment of liquid water, soft-cutoff conditioning doesn't  change 
the fundamental convergence characteristics of the hard-cutoff approach.
Hard cutoffs are found to work better than softer ones in that case.

\section{Acknowledgment}
This work was carried out under the auspices of the National Nuclear
Security Administration of the U.S. Department of Energy at Los Alamos
National Laboratory under Contract No. DE-AC52-06NA25396. 

\vfill
\newpage


\bibliographystyle{jpc4}

\end{document}